\documentclass[12pt]{iopart}
\pdfoutput=1
\usepackage{iopams}
\expandafter\let\csname equation*\endcsname\relax
\expandafter\let\csname endequation*\endcsname\relax
\usepackage{amsmath}
\usepackage{graphicx}
\usepackage{color}
\usepackage{nicefrac}

\begin{document}


\title{Hierarchical Bounds on Entropy Production Inferred from Partial Information}

\author{Gili Bisker$^1$, Matteo Polettini$^2$, Todd R. Gingrich$^1$ and Jordan M. Horowitz$^1$}

\address{$^1$ Physics of Living Systems Group, Department of Physics, Massachusetts Institute of Technology, 400 Technology Square, Cambridge, MA 02139}
\address{$^2$ Complex Systems and Statistical Mechanics, Physics and Materials Research Unit, University of Luxembourg,
162a Avenue de la Fa\"{i}encerie, L-1511 Luxembourg, (G. D. Luxembourg)}
\ead{bisker@mit.edu}

\date{\today}

\begin{abstract}
Systems driven away from thermal equilibrium constantly deliver entropy to their environment.
Determining this entropy production requires detailed information about the system's internal states and dynamics.
However, in most practical scenarios, only a part of a complex experimental system is accessible to an external observer.
In order to address this challenge, two notions of partial entropy production have been introduced in the literature as a way to assign an entropy production to an observed subsystem: one due to Shiraishi and Sagawa [{\it Phys. Rev. E} {\bf 91}, 012130 (2015)] and another due to Polettini and Esposito [arXiv:1703.05715 (2017)].
We show that although both of these schemes {provide} a lower bound on the total entropy production, the latter -- which utilizes an effective thermodynamics description-- gives a better estimate of the total dissipation. 
Using this effective thermodynamic framework, we establish a partitioning of the total entropy production into two contributions that individually verify integral fluctuation theorems: an observable partial entropy production and a hidden entropy production assigned to the unobserved subsystem. 
Our results offer broad implications for both theoretical and empirical systems when only partial information is available.

\end{abstract}



\maketitle


\section{Introduction}

Stochastic thermodynamics has refined our understanding of dissipation at the mesoscale by unraveling the thermodynamic content of  fluctuations~\cite{Jarzynski2011,Seifert2012}.
As the dissipation and its fluctuations are a central object of the theory, their calculation and measurement is paramount.
However, determining the total dissipation requires one to carefully track in full detail a system's mesoscopic dynamics, which may not always be possible: 
experiments may only be able to resolve a subset of the degrees of freedom~\cite{Mehl2012, Chiang2016}, or calculations may be impractical for systems with many internal states~\cite{Frenkel2002}.
Thus, a consistent approach for treating the fluctuating entropy production $\sigma$ with only partial information is a necessary aspect for any useful nonequilibrium thermodynamic framework. 

One could imagine two notions of partial information.
The first utilizes coarse graining, where several states are clumped together or traced out; thereby, obscuring any internal dissipation.
Such a framework has been studied extensively from the point of view of stochastic thermodynamics, both theoretically~\cite{Kawai2007,Kawaguchi2013,Gomez-Marin2008b,Rahav2007,Esposito2012c,Altaner2012,Zimmermann2015,Borrelli2015,Horowitz2009b,Esposito2014,Puglisi2010,Celani2012} and experimentally~\cite{Mehl2012, Chiang2016}. 
The second notion, and the one we consider here, is that the observer has access to only a subset of system states; the rest are hidden or masked.
Having this point of view, clearly distinguishes between the observed part of the system and its hidden counterpart, inviting the challenge of decomposing the total entropy production into partial entropy productions for both subsystems.

When the observer only has access to a subset of states, two approaches to assigning fluctuating partial entropy production $\sigma^{\rm part}$ have been introduced in the literature, both of which verify fluctuation theorems.  
The first, due to Shiraishi and Sagawa~\cite{Shiraishi2014,Shiraishi2015,Shiraishi2016} (see also~\cite{Rosinberg2016}), was developed in part to provide a fluctuating counterpart to the thermodynamics of continuous information flow~\cite{Parrondo2015,Horowitz2014,horowitz2014b,Horowitz2015,Barato2014b}.
A similar construction was also proposed by Hartich, Barato and Seifert~\cite{Hartich2014} in the context of bipartite systems. 
The distinguishing features here are that the partial system entropy is inferred from a passive observation of a subsystem
and that the true thermodynamic force is utilized.
As such, we will  refer to this approach as the passive partial entropy production to emphasize that the observer does not need to manipulate the system in this framework.
By contrast, Polettini and Esposito recently suggested an alternative approach for assigning partial entropy production, which incorporates an effective thermodynamic force at the cost of demanding that the observer has control over the observed dynamics \cite{Polettini2017}.
As this version requires additional information  regarding the effect of external control parameters on the dynamics, we refer to this construction as the informed partial entropy production.

\begin{figure}[t]
\begin{center}
\includegraphics[scale=0.6]{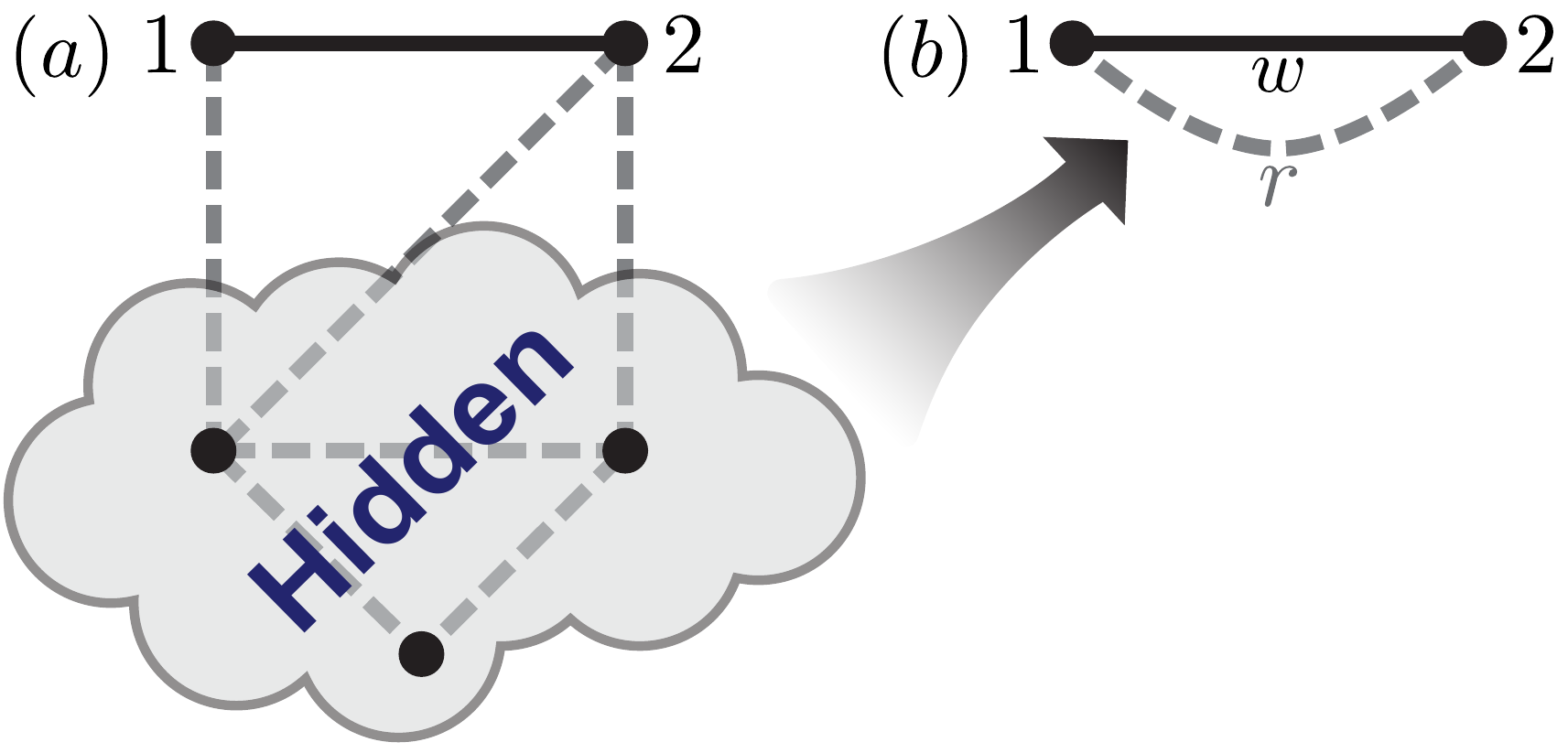}
\caption{\label{setup}Illustration of partially observed thermodynamics: \emph{(a)} An observer can measure the currents and probabilities for the $1-2$ link, whereas the rest of the system is hidden. \emph{(b)} The observer can assign an effective description to the hidden part by coarse-graining the hidden network to one effective transition with rates $r$.}
\end{center}
\end{figure}

In this article, we discuss both the passive and informed partial entropy production approaches from a unifying perspective, provide insights and intuition, as well as  extend the current understanding of these frameworks.
First, we show that both partial entropy productions naturally lead to a decomposition of the total dissipation  $\sigma$ into two positive fluctuating pieces, as
\begin{equation}
\sigma=\sigma^{\rm part}+\sigma^{\rm comp},
\end{equation}
where each contribution -- the partial entropy production $\sigma^{\rm part}$ and its complement $\sigma^{\rm comp}$ -- individually satisfy an integral fluctuation theorem
\begin{equation}
\langle e^{-\sigma^{\rm part}}\rangle = 1, \qquad \langle e^{-\sigma^{\rm comp}}\rangle=1,
\end{equation}
and as such are individually positive  \cite{Esposito2010},
\begin{equation}
\langle \sigma^{\rm part}\rangle\ge 1, \qquad \langle \sigma^{\rm comp}\rangle\ge 1.
\end{equation}
Shiraishi and Sagawa proved these relationships quite generally for  the passive partial entropy production \cite{Shiraishi2014}, whereas here, we develop this decomposition for the informed partial entropy production of Polettini and Esposito, both for stationary nonequilibrium steady states as well as an extended version for transient driven dynamics. 
With these tools in hand, we then show that owing to the extra physical information incorporated in the informed partial entropy production, it is always larger on average than the passive partial entropy production, demonstrating a precise hierarchy in partial measures of entropy production.

The manuscript is organized as follows: In \sref{sec:setup}, we lay the foundations for our model system, which is a continuous-time Markov jump process on a network of mesoscopic states, as well as a general derivation of the fluctuation theorem for the total entropy production. We then introduce the notion of partial entropy production in \sref{sec:Partial entropy production} and discuss the two approaches. Subsequently, we derive the partial entropy production for the hidden part of the dynamics in \sref{sec:Entropy production decomposition}, where we demonstrate that the total entropy production can be decomposed into two positive contributions corresponding to the observed and hidden parts. In  \sref{sec:Entropy production hierarchy}, we compare the passive and informed partial entropy productions to prove 
the hierarchical order between them. 
The advantage of the informed partial entropy production framework is further demonstrated in \sref{sec:Partial information can be complete}, where we show that for a unicycle network it reproduces the total entropy production exactly. 
As a final bit of analysis, we extend the informed partial entropy production approach to time-dependent {driven} dynamics in \sref{sec:Time-dependent partial entropy production}. 
In \sref{sec:Numerical simulations}, we present a numerical case study to illustrate our main results, and we conclude with a thorough discussion and outlook in \sref{sec:Discussion}. Supplementary calculations can be found in the Appendices.

\section{Setup}\label{sec:setup}

We begin our analysis by first describing the dynamics and thermodynamics of our system of interest.
With the context fixed, we then introduce fluctuation theorems from a general perspective as symmetries of trajectory observables obtained from logratios of trajectory probabilities.
This will set the stage for our comparison of partial entropy productions as trajectory observables.

\subsection{Model system}
We consider a mesoscopic system modeled as a continuous-time Markov jump process over a finite set of states $\{1,...,K\}$. 
The probability density ${\bf p}(t)=\{p_i(t)\}$ then obeys the Master Equation 
\begin{align}
{\dot {\bf p}}(t)&= {\bf W}{\bf p}(t),
\end{align}
where the transition rate matrix
\begin{equation}
W_{ij}=\left\{
\begin{array}{ll}
w_{ij} & i\neq j \\
-\lambda_{i} & i=j \\
\end{array}\right. ,
\end{equation}
encodes the rates $w_{ij}$ to jump from $j\to i$ on the off-diagonal elements and the escape rates $\lambda_{j}=\sum_{i\neq j}w_{ij}$ on the diagonal elements, which enforce probability conservation.
As such, we can identify the (probability) current flowing from $j\to i$ as
\begin{equation}
J_{ij}(t)=w_{ij}p_j(t)-w_{ji}p_i(t).
\end{equation}
We assume that each transition is reversible, that is $w_{ij}>0$ only when $w_{ji}>0$, and that there is a unique stationary state ${\boldsymbol \pi}=\{\pi_i\}$ satisfying ${\bf W}\boldsymbol\pi=0$, with stationary current $J^\pi_{ij}=w_{ij}\pi_j-w_{ji}\pi_i$.

For a thermodynamically consistent description, we assume that local detailed balance holds, so that every transition is accompanied by a fixed entropy flow into the environment.
The second law of thermodynamics then dictates that the steady-state entropy production rate is positive~\cite{Seifert2012,Esposito2010b}
\begin{equation}\label{eq:avgEP}
\Sigma=\sum_{i<j}J^\pi_{ij}\ln\frac{w_{ij}\pi_j}{w_{ji}\pi_i}\equiv\sum_{i<j}J^\pi_{ij}F_{ij}\ge 0,
\end{equation}
which defines the steady-state thermodynamic force, or affinity, $F$ that measures the entropy flow into the thermal reservoir mediating the transition.
Clearly, observing this entropy production requires one to be able to monitor every transition in order to determine every term in the sum.
The partial entropy productions that we discuss, however, circumvent this requirement.
In order to lay the foundations for this framework, let us now turn to fluctuation theorems and their relation to entropy production.

\subsection{Fluctuation theorems from auxiliary dynamics}

Fluctuation theorems deal with symmetries of certain trajectory observables and are generically derived by comparing the probability to observe a mesoscopic trajectory and its time reverse in a possibly distinct auxiliary dynamics~\cite{Harris2007,Chetrite2008,Esposito2010,Lacoste2012,Baiesi2015,Shargel2010}.
The great freedom in this construction, which has led to the proliferation of fluctuation theorems, is that we may choose any generator ${{\bf\bar W}}$ for the auxiliary Markov process.
Some choices turn out to have clear and interesting physical interpretations, such as the two that give rise to the partial entropy productions, which are the focus of this paper.

Specifically, for a fixed observation time $T$, let us denote a trajectory by $\gamma=\{(i_0,\tau_0),\dots, (i_N,\tau_N)\}$ -- which is a chronological sequence of the $N$ states $\{i_0,\dots,i_N\}$ visited during the trajectory and their wait times $\{\tau_0,\dots,\tau_N\}$, with $\sum_i\tau_i=T$ -- and its time reverse by $\tilde\gamma=\{(i_N,\tau_N),\dots, (i_0,\tau_0)\}$.
The probability ${\mathcal P}[\gamma]$ of observing $\gamma$ is thus~\cite{Seifert2012}
\begin{equation}
{\mathcal P}[\gamma]=e^{-\tau_N\lambda_{ i_N}}\prod_{n=0}^{N-1}\left[w_{i_{n+1},i_n}e^{-\tau_n\lambda_{i_n}}\right]\pi_{i_0},
\end{equation}
{where the initial state is sampled from the steady state distribution ${\boldsymbol \pi}$.}
Then we can construct a trajectory observable from the ratio of ${\mathcal P}[\gamma]$ and the probability $\bar{\mathcal P}[\tilde\gamma]$ of observing the reverse trajectory $\tilde\gamma$ in an auxiliary dynamics~\cite{Lacoste2012},
\begin{equation}\label{eq:trajObs}
\begin{split}
{\mathcal R}[\gamma]&=\ln\frac{{\mathcal P}[\gamma]}{\bar{\mathcal P}[\tilde\gamma]}\\
&=\ln\frac{\pi_{i_0}}{ \pi_{ i_N}}+\sum_{n=0}^{N-1}\ln\frac{w_{i_{n+1},i_n}}{ \bar w_{ i_{n}, i_{n+1}}} - \sum_{n=0}^{N}(\lambda_{i_n}-\bar\lambda_{ i_n})\tau_n
\end{split}
\end{equation}
Being a logratio of probabilities, ${\mathcal R}$ immediately satisfies an integral fluctuation theorem $\langle e^{-{\mathcal R}}\rangle=1$, as can be easily checked \cite{Esposito2010}.

A particularly important example of a trajectory observable is the fluctuating steady-state entropy production~\cite{Seifert2012}
\begin{equation}\label{eq:totEP}
\begin{split}
\sigma&= \ln\frac{\pi_{i_0}}{\pi_{i_N}} + \sum_{n=0}^{N-1} \ln \frac{w_{i_{n+1},i_n}}{w_{i_{n},i_{n+1}}}\\
&=\ln\frac{\pi_{i_0}}{\pi_{i_N}} + \sum_{i < j}\phi_{ij}\ln \frac{w_{ij}}{w_{ji}},
\end{split}
\end{equation}
with long-time average $\Sigma=\lim_{T\to\infty}\langle\sigma\rangle/T$, and where $\phi_{ij}$ is the net number of transitions from $j\to i$ over the course of the trajectory $\gamma$: 
\begin{equation}
\phi_{ij}=\sum_{n=0}^{N-1}\left(\delta_{i,i_{n+1}}\delta_{j,i_n} - \delta_{j,i_{n+1}}\delta_{i,i_n}   \right).
\end{equation}
Here, the auxiliary generator is simply the same as the original: plugging ${{\bf\bar W}} = \bf W$ into \eqref{eq:trajObs} leads to the total entropy production $\sigma$ in \eqref{eq:totEP}: $\sigma=\ln({\mathcal P}[\gamma]/{\mathcal P}[\tilde\gamma])$.

An alternative formulation that will shed light on our discussion of partial entropy productions is to utilize a special auxiliary dynamics called the dual process whose generator implements time-reversal~\cite{Crooks1999a,Crooks2000},
\begin{equation}\label{eq:revGen}
{\bar W}^{\rm dual}_{ij}=\left\{
\begin{array}{ll}
w_{ji}\frac{\pi_i}{\pi_j} & i\neq j \\
-\lambda_{i} & i=j \\
\end{array}\right.,
\end{equation}
which ``twists'' all the transition rates with a weight $\pi_i/\pi_j$.
These dynamics have the special property that they generate the reverse trajectories with the same probabilities as the original process: $\bar{\mathcal P}^{\rm dual}[\gamma]={\mathcal P}[\tilde\gamma]$.
As such, the total entropy production {can be alternatively derived as $\sigma=\ln({\mathcal P}[\gamma]/{\mathcal P}[\tilde\gamma])=\ln({\mathcal P}[\gamma]/\bar{\mathcal P}^{\rm dual}[\gamma])$}.

\section{Partial entropy production}\label{sec:Partial entropy production}

Calculating the total entropy production, according to \eqref{eq:totEP}, requires complete knowledge of the system dynamics;  
an external observer needs to record every step of a trajectory. 
However, all this information is not always readily available, requiring the development of partial entropy productions.

In this section, we compare and contrast two fluctuating partial entropy productions both of which satisfy integral fluctuation theorems.
To keep the discussion as concrete as possible, we specialize to a system at steady state, where the observer can only monitor two states, $1$ and $2$, and transitions between them (Figure \ref{setup}).
In particular, they can only measure (or calculate), the steady state probabilities of the observed states, $\pi_1$ and $\pi_2$, and the average rate of jumps between them, $w_{21}$ and $w_{12}$.

The key insight that allows the development of the fluctuation theorems for both partial entropy productions, turns out also to be the unifying perspective.
Both partial entropy productions are trajectory observables where the auxiliary generator is obtained by twisting a subset of the transitions; namely, the hidden transitions~\cite{Shiraishi2014,Polettini2017}
\begin{equation}\label{auxiliary process for single link}
{\bar W}_{ij}=\left\{
\begin{array}{ll}
w_{ij} & ij=12,21\\
w_{ji}\frac{u_i}{u_j} & i\neq j; \, ij\neq12,21 \\
-{\bar \lambda}_{i} & i=j \\
\end{array}\right.,
\end{equation}
with each $u_i>0$ and the ${\bar\lambda_i}$ chosen to enforce probability conservation.
As we will see, the choice of $u$ determines the partial entropy production.

\subsection{Passive partial entropy production}

The passive partial entropy production $\sigma^{\rm  pp}$ identified by Shiraishi and Sagawa \cite{Shiraishi2014} takes the form in our restricted setup
\begin{equation}\label{single link EP}
\sigma^{\rm  pp}=\phi_{12}\ln\frac{w_{12}\pi_{2}}{w_{21}\pi_{1}}-\left(J^\pi_{12}\frac{T_1}{\pi_1}+J^\pi_{21}\frac{T_2}{\pi_2} \right),
\end{equation}
where $T_j$ is the total fluctuating time spent in state $j$ over the course of a trajectory.
This definition should be compared with that introduced by Hartich, Barato and Seifert~\cite{Hartich2014}, which has a plus sign in front of the parenthesis.
The physical significance of \eqref{single link EP} is most apparent if we look at the average entropy production rate in the steady-state limit
\begin{equation}
\Sigma^{\rm pp}=\lim_{T\to\infty}\frac{1}{T}\langle \sigma^{\rm pp}\rangle=J^\pi_{12}\ln\frac{w_{12}\pi_{2}}{w_{21}\pi_{1}}\ge 0,
\end{equation}
where we have used the ergodicity assumption that within this limit 
{${T_j}/{T}$ converges to $\pi_j$, and ${\phi_{ij}}/{T}$ converges to $J^\pi_{ij}$.}
Upon comparison with the average total entropy production \eqref{eq:avgEP}, we see this is simply the contribution coming just from transitions between states $1$ and $2$; a natural choice for the partial entropy production.

The fluctuation theorem for \eqref{single link EP} arises from an auxiliary process where  the twisting parameters are simply the steady-state probabilities, $u_i=\pi_i$~\cite{Shiraishi2014}:
\begin{equation}\label{eq:ppDual}
{\bar W}^{\rm pp}_{ij}=\left\{
\begin{array}{ll}
w_{ij} & ij=12,21\\
w_{ji}\frac{\pi_i}{\pi_j} & i\neq j; \, ij\neq12,21 \\
-{\bar \lambda}_{i} & i=j \\
\end{array}\right.,
\end{equation}
with modified exit rates {that guarantee conservation of probability}
\begin{equation}\label{single_link_duel_exit_time}
{\bar \lambda_{i}}=\left\{ 
\begin{array}{ll}
\lambda_{i}+\frac{1}{\pi_i}\sum_{j\neq 1,2}J^\pi_{ij} & i=1,2 \\
\lambda_{i} & i\neq1,2 \\
\end{array}\right..
\end{equation}
See \ref{full_derivation_SL} for a detailed derivation. In essence, the twisting generates the reverse dynamics (cf.~\eqref{eq:revGen}) on the hidden states.

\subsection{Informed partial entropy production}

The informed partial entropy production requires an additional assumption~\cite{Polettini2017}, that the observer can tune the observed transition rates, $w_{12}(x)$, and $w_{21}(x)$, by varying an external control parameter or force $x$.
As we will see, this additional freedom allows one to identify and measure an alternative notion of partial entropy production.

Let us denote the parameter-dependent generator as ${\bf W}(x)$, which is assumed to have a unique steady state distribution $\boldsymbol\pi(x)$ for every value of $x$.
Now, the informed partial entropy production is based on the observation that there is a special value of the control parameter where the steady-state current on the $1-2$ transition is zero, which we call the stalling force $x^{\rm st}$: $w_{12}(x^{\rm st})\pi_2(x^{\rm st}) - w_{21}(x^{\rm st})\pi_1(x^{\rm st})=0$.
This lack of current immediately connects the stalling steady-state distribution to the transition rates in a simple way:
\begin{equation}\label{eq:stalling rates and rates}
\frac{w_{12}(x^{\rm st})}{w_{21}(x^{\rm st})}=\frac{\pi_1(x^{\rm st})}{\pi_2(x^{\rm st})}\equiv\frac{\pi^{\rm st}_1}{\pi^{\rm st}_2}.
\end{equation}
Mathematically, the distribution $\boldsymbol\pi^{\rm st}$ can be obtained as the steady-state of a modified generator $\bf W^{\rm st}$ with the $1-2$ transitions removed:  ${\bf{W}}^{\rm st}{\boldsymbol\pi}^{\rm st}=0$.
This is apparent, since $\boldsymbol\pi^{\rm st}$ represents the steady-state with vanishing current (no net transitions $1\leftrightarrow 2$), which can be enforced simply by setting $w_{12}=w_{21}=0$, as illustrated in Figure \ref{stalling}. 
Details are in \ref{Appendix:Stalling distribution}.

\begin{figure}
\begin{center}
\includegraphics[scale=0.6]{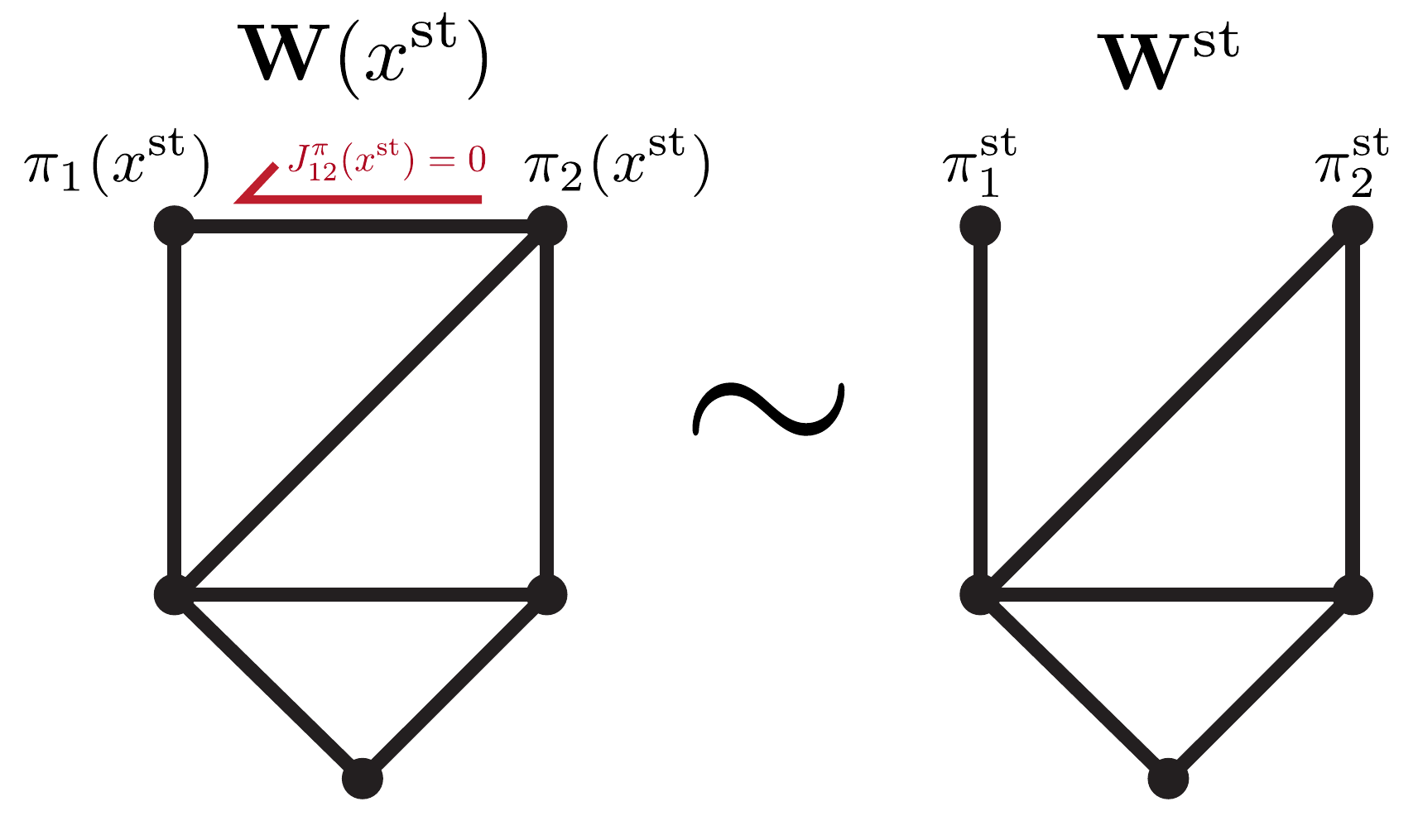}
\caption{\label{stalling} Illustration of the stalling distribution: At the stalling force, the current over the observed $1-2$ link vanishes, leading to a stalling steady-state distribution ${\bf \pi}(x^{\rm st})$ (left). This situation is analogous to having zero rates on the observed link (\emph{i.e.}~removing it completely), leading to the same steady-state distribution (right).}
\end{center}
\end{figure}

Now, the informed partial entropy production {$\sigma^{\rm ip}$} (for any value of $x$) is defined in a manner akin to \eqref{single link EP}, except using the stalling distribution~\cite{Polettini2017}, here extended to transient trajectories, 
\begin{equation}\label{FT_M}
\sigma^{\rm ip}=\ln\frac{\pi_{i_0}\pi^{\rm st}_{i_N}}{\pi_{i_N}\pi^{\rm st}_{i_{0}}}+\phi_{12}\ln\frac{w_{12}\pi^{\rm st}_{2}}{w_{21}\pi^{\rm st}_{1}},
\end{equation}
with average steady-state rate
\begin{equation}
\Sigma^{\rm ip}=\lim_{T\to\infty}\frac{1}{T}\langle \sigma^{\rm ip}\rangle=J^\pi_{12}\ln\frac{w_{12}\pi^{\rm st}_{2}}{w_{21}\pi^{\rm st}_{1}}.
\end{equation}
The rational behind this definition has a profound physical significance.
This entropy production is predicated on an effective thermodynamic description of the system as perceived by the observer.
In effect, the observer sees a nontrivial effective thermodynamic force~\cite{Polettini2017} 
\begin{equation}\label{eq: def of r rates}
F^{\rm st}=\ln\frac{w_{12}\pi^{\rm st}_2}{{w_{21}}\pi^{\rm st}_1}.
\end{equation}
This description is consistent with a minimal model that captures the observed steady-state dynamics by collapsing the hidden part of the network to a single transition with \emph{parameter-independent} rates  (as depicted in Figure \ref{setup}b):
\begin{equation}\label{eq:rRatio}
\frac{r_{21}}{r_{12}}=\frac{\pi^{\rm st}_2}{\pi^{\rm st}_1}
\end{equation}
which are defined to maintain the correct steady-state density for every parameter value:
\begin{equation}\label{def_w_tilde}
\frac{w_{12}(x)+r_{12}}{w_{21}(x)+r_{21}}=\frac{\pi_1(x)}{\pi_2(x)}.
\end{equation}
Importantly, the rates $r$ are uniquely defined and can be determined from $\bf W^{\rm st}$, independent of $x$ (see \ref{Appendix:Stalling distribution})~\cite{Polettini2017}.

Underlying the identification of \eqref{FT_M} as an entropy production is an integral fluctuation theorem.
Here we choose the twisting parameters to  be the stalling distribution, $u_i=\pi^{\rm st}_i$~\cite{Polettini2017}:
\begin{equation}\label{eq:def of Wip}
{\bar W}^{\rm ip}_{ij}=\left\{
\begin{array}{ll}
w_{ij} & ij=12,21\\
w_{ji}\frac{\pi^{\rm st}_i}{\pi^{\rm st}_j} & i\neq j; \, ij\neq12,21 \\
-\lambda_{i} & i=j \\
\end{array}\right.,
\end{equation}
where remarkably the exit rates $\lambda_i$ are unmodified (See \ref{full_derivation_SL} for details).
In fact, this property singles out the twisting $u_i=\pi^{\rm st}_i$ as unique.

\subsection{Summary}

Ultimately, the formal structure of the two partial entropy productions are the same.
Both verify integral fluctuation theorems obtained by twisting the generator on the hidden network with a normalized probability distribution.
However, the physical significance of the two entropy productions are distinct, owing to the two different choices of twistings.
In the following, we will explore their relationship.

\section{Entropy production decomposition}\label{sec:Entropy production decomposition}

So far, we have laid out the two different approaches for assigning entropy production to a single observable link and the corresponding fluctuation theorems.  
Further insight into their comparison comes from analyzing the complementary entropy production in the hidden part of the network.

\subsection{Passive partial entropy production}
According to \cite{Shiraishi2014}, the hidden part of the entropy production, $\sigma^{\rm pp,c}\equiv \sigma-\sigma^{\rm pp}$ -- with ``$\rm c$'' standing for complement -- satisfies a fluctuation theorem.
Meaning, it can be written as the logratio between two trajectory probability distributions. 
However, one has to define a new auxiliary process analogously to the definition in \eqref{eq:ppDual}, except treating the $1-2$ link as hidden~\cite{Shiraishi2014}:
\begin{equation}\label{hidden auxiliary process for single link}
{\bar W}_{ij}^{\rm pp,{c}}=\left\{
\begin{array}{ll}
w_{ji}\frac{\pi_i}{\pi_j} &   ij=12,21\\
w_{ij} & i\neq j, \quad ij\neq12,21\\
-{\bar \lambda}_{i}^{\rm {c}} & i=j \\
\end{array}\right.,
\end{equation}
with modified exit rates chosen to conserve probability,
\begin{equation}\label{hidden single_link_duel_exit_time}
{\bar \lambda}_{i}^{\rm c}=\left\{ 
\begin{array}{ll}
\lambda_{1}+\frac{1}{\pi_1}J^\pi_{12} & i=1 \\
\lambda_{2}+\frac{1}{\pi_2}J^\pi_{21} & i=2 \\
\lambda_{i} & i\neq1,2 \\
\end{array}\right. .
\end{equation}
This construction naturally leads to a trajectory observable (cf. Eq.~\eqref{eq:trajObs})
\begin{equation}\label{eq:EPcomp}
\begin{split}
&\sigma^{\rm pp,c}=\ln\frac{\mathcal{P}[\gamma]}{ \bar{\mathcal P}^{\rm pp,c}[\tilde\gamma]}\\
&=\sum_{\substack{i<j\\(i,j)\neq(1,2)}}\phi_{ij}\ln\frac{w_{ij}\pi_j}{ w_{ji}\pi_i}+\left(J^{\pi}_{12}\frac{T_1}{\pi_1}+J^{\pi}_{21}\frac{T_2}{\pi_2} \right),
\end{split}
\end{equation}
with average rate
\begin{equation}
\Sigma^{\rm pp,c}=\lim_{T\to\infty}\frac{1}{T}\langle\sigma^{\rm pp,c}\rangle=\sum_{\substack {i<j \\ (i,j)\neq(1,2)}}J^\pi_{ij}\ln\frac{w_{ij}\pi_j}{ w_{ji}\pi_i}.
\end{equation}
Thus, this complementary entropy production is simply the entropy production arising from all the hidden transitions. 
See  \ref{full_derivation_SM_unobserved} for a derivation.

From their trajectory definitions, \eqref{single link EP} and \eqref{eq:EPcomp}, it is straightforward to check that indeed (see  \ref{full_derivation_decomposition}):
\begin{equation}\label{decomposition shiraishi}
\sigma=\sigma^{\rm pp}+\sigma^{\rm pp,c}.
\end{equation}
The fact that such a decomposition exists is perhaps more surprising, when we reframe this equation using logratios of trajectory distributions
\begin{equation}
\underbrace{\ln\frac{\mathcal{P}[\gamma]}{{\mathcal P}[\tilde\gamma]}}_{\sigma}=\underbrace{\ln\frac{\mathcal{P}[\gamma]}{ \bar{\mathcal P}^{\rm pp}[\tilde\gamma]}}_{\sigma^{\rm pp}}+\underbrace{\ln\frac{\mathcal{P}[\gamma]}{ \bar{\mathcal P}^{\rm pp,c}[\tilde\gamma]}}_{\sigma^{\rm pp,c}}.
\end{equation}
This decomposition requires the conclusion that the auxiliary processes verify
\begin{equation}\label{eq:ppPratio}
\frac{\mathcal{P}[\gamma]}{ \bar{\mathcal P}^{\rm pp,c}[\tilde\gamma]}=\frac{\bar{\mathcal P}^{\rm pp}[\tilde\gamma]}{ {\mathcal P}[\tilde\gamma]}.
\end{equation} 
Meaning, the hidden auxiliary process interchanges the ratio of distributions; a rather unique time-reversal-like structure.

\subsection{Informed partial entropy production}

Polettini and Esposito did not derive a complementary entropy production in their original work~\cite{Polettini2017}.
Such a decomposition though is possible, as we show in this section, which constitutes our first main result.

Remarkably, the situation is much simpler here as we do not need to define a new auxiliary process.
Instead, the complementary informed partial entropy production can be deduced by considering
\begin{equation}
\sigma^{\rm ip,c}=\ln\frac{{\mathcal P}[\gamma]}{{\bar{\mathcal P}}^{\rm ip}[\gamma]} =\sum_{n=0}^{N-1}\ln\frac{w_{i_{n+1},i_n}}{ {\bar w}_{i_{n+1},i_n}},
\end{equation}
where both trajectory distributions are evaluated on the {{\it same}} trajectory.
Since the rates over the $1-2$ link are unaltered in the auxiliary generator ${\bf W}^{\rm ip}$ (cf. Eq.~\eqref{eq:def of Wip}), the only contributions to the sum are from jumps over the hidden transition:
 \begin{equation}
 \sigma^{\rm ip,c}=\ln\frac{{\mathcal P}[\gamma]}{{\bar {\mathcal P}}^{\rm ip}[\gamma]} = \sum_{\substack{i<j\\(i,j)\neq(1,2)}}\phi_{ij}\ln\frac{w_{ij}\pi^{\rm st}_{j}}{ w_{ji}\pi^{\rm st}_{i}}.
 \end{equation}
Summing up the contributions of the observed and hidden parts, it is straightforward to verify that (see  \ref{full_derivation_decomposition}):
\begin{equation}\label{decomposition matteo}
\sigma=\sigma^{\rm ip}+\sigma^{\rm ip,c}.
\end{equation}

In terms of the trajectory distributions, this decomposition rests on the remarkable property of the auxiliary process
\begin{equation}\label{eq:ipPratio}
\frac{{\mathcal P}[\gamma]}{\bar{\mathcal P}^{\rm ip}[\gamma]}=\frac{\bar{\mathcal P}^{\rm ip}[\tilde\gamma]}{{\mathcal P}[\tilde\gamma]}.
\end{equation}
Time reversing flips the ratio of probabilities.
The essential feature that allows for such a unique property (and decomposition) is the fact that the escape rates are unaltered in the auxiliary dynamics.
As pointed out in \cite{Lacoste2012}, and manifested in \eqref{eq:trajObs}, ratios between trajectory probabilities generated from two distinct dynamics include terms like the last term in \eqref{eq:trajObs}  corresponding to the difference in escape rates, or the traffic, between the dynamics.
This is precisely the source of the expression in parenthesis in the definition of $\sigma^{\rm pp}$ in \eqref{single link EP}, which depends on {waiting} times.
Generically, this term hinders a simple and elegant decomposition using a single auxiliary process. 
The different auxiliary process we had to introduce for the hidden part of the passive partial entropy production \eqref{hidden auxiliary process for single link}, also had different escape rates with respect to the original dynamics.
The underlying reason was so that the traffic terms in the definitions of $\sigma^{\rm pp,c}$ and $\sigma^{\rm pp}$ canceled, rendering their sum the total entropy production.
In contrast, for the informed partial entropy production, the same auxiliary process was used to recover both the observed and hidden parts of the total entropy production, neither of which included a traffic term. 
As pointed out in the previous section, this feature distinguishes the definition of the auxiliary process for the informed partial entropy production.

Let us note that a similar utilization of a single auxiliary process with escape rates identical to the original dynamics was employed to decompose the total entropy production for driven dynamics into adiabatic and non-adiabatic parts~\cite{Esposito2010}. 
There too, the decomposition was facilitated by the fact that the trajectory probability ratios did not include contributions from differences in the diagonal elements of the generator matrices.

\subsection{Summary}

We emphasize that we have two decompositions of the total entropy production into a pair of positive (on average) parts that each verify an integral fluctuation theorem.
Underlying these decompositions are a pair of auxiliary processes that share special symmetry properties with the original dynamics under time-reversal.
However, the informed partial entropy production is singled out by the property that its auxiliary generator maintains the escape rates, implying the partial entropy production and its complement can be constructed from ratios of a pair of trajectories, either forward or reverse, generated from the same dynamics.

One consequence of this profusion of entropic measures is that we now have four distinct lower bounds on the average entropy production $\langle \sigma\rangle$: 
\begin{equation}
\langle \sigma\rangle \ge \{\langle\sigma^{\rm pp}\rangle,\ \langle\sigma^{\rm pp,c}\rangle,\ \langle\sigma^{\rm ip}\rangle,\ \langle\sigma^{\rm ip,c}\rangle\}.
\end{equation}
In the following section, we rationalize this structure, by demonstrating a hierarchy of entropy productions.

\section{Entropy production hierarchy}\label{sec:Entropy production hierarchy}

The partial entropy productions assigned to a single observed link both satisfy integral fluctuation theorems and provide a lower bound on the total entropy production. 
In this section, we compare these two expressions, showing that the informed partial entropy production is always greater owing to the additional physical information incorporated in its definition.
We will focus on the average entropy production rates for both cases, $\Sigma^{\rm ip}$ and $\Sigma^{\rm pp}$, which dominate in the long time limit.

To deduce an inequality between $\Sigma^{\rm ip}$ and $\Sigma^{\rm pp}$, we consider their difference 
\begin{equation}\label{eq:diff of EP rates}
\Sigma^{\rm ip}-\Sigma^{\rm pp}=(w_{12}\pi_2 - w_{21}\pi_1)\ln\frac{\pi^{\rm st}_2 \pi_1}{\pi^{\rm st}_1 \pi_2}.
\end{equation}
According to (\ref{def_w_tilde}), we have that
\begin{equation}\label{eq:eq of currents}
w_{12}\pi_2 - w_{21}\pi_1 =   r_{21}\pi_1 - r_{12}\pi_2. 
\end{equation}
Substituting \eqref{eq:rRatio} and \eqref{eq:eq of currents} into \eqref{eq:diff of EP rates} leads to
\begin{equation}
\Sigma^{\rm ip}-\Sigma^{\rm pp}=(r_{12}\pi_2 - r_{21}\pi_1)\ln\frac{r_{12} \pi_2}{r_{21} \pi_1}\ge 0,
\end{equation}
with positivity due to the convexity of the logarithm, $(x-y)\ln(x/y)\ge0$.

Combined with our previous results, we have an entropy production hierarchy:
\begin{equation}\label{hierarchy}
\langle \sigma \rangle \ge \langle \sigma^{\rm ip}\rangle \ge \langle\sigma^{\rm pp}\rangle \ge 0,
\end{equation}
which is our second main result. 
Consequently, $\langle \sigma^{\rm ip}\rangle$ offers a better estimate of the total dissipation in the system whenever only partial information is available.
However, determining $\langle\sigma^{\rm ip}\rangle$ requires additional input as compared to $\langle \sigma^{\rm pp}\rangle$; namely, knowledge of the stationary stalling probabilities of the two observed states.
We stress that the stalling distribution can be obtained by manipulating  the observed transition solely, without having to affect the hidden part of the network. Tuning the rates of the observed link in order to find the stalling probabilities might not always be readily attainable.
However, in situations where it is possible, there is a true gain in obtaining these data.

\section{Partial information can be complete}\label{sec:Partial information can be complete}
\begin{figure}
\begin{center}
\includegraphics[scale=0.6]{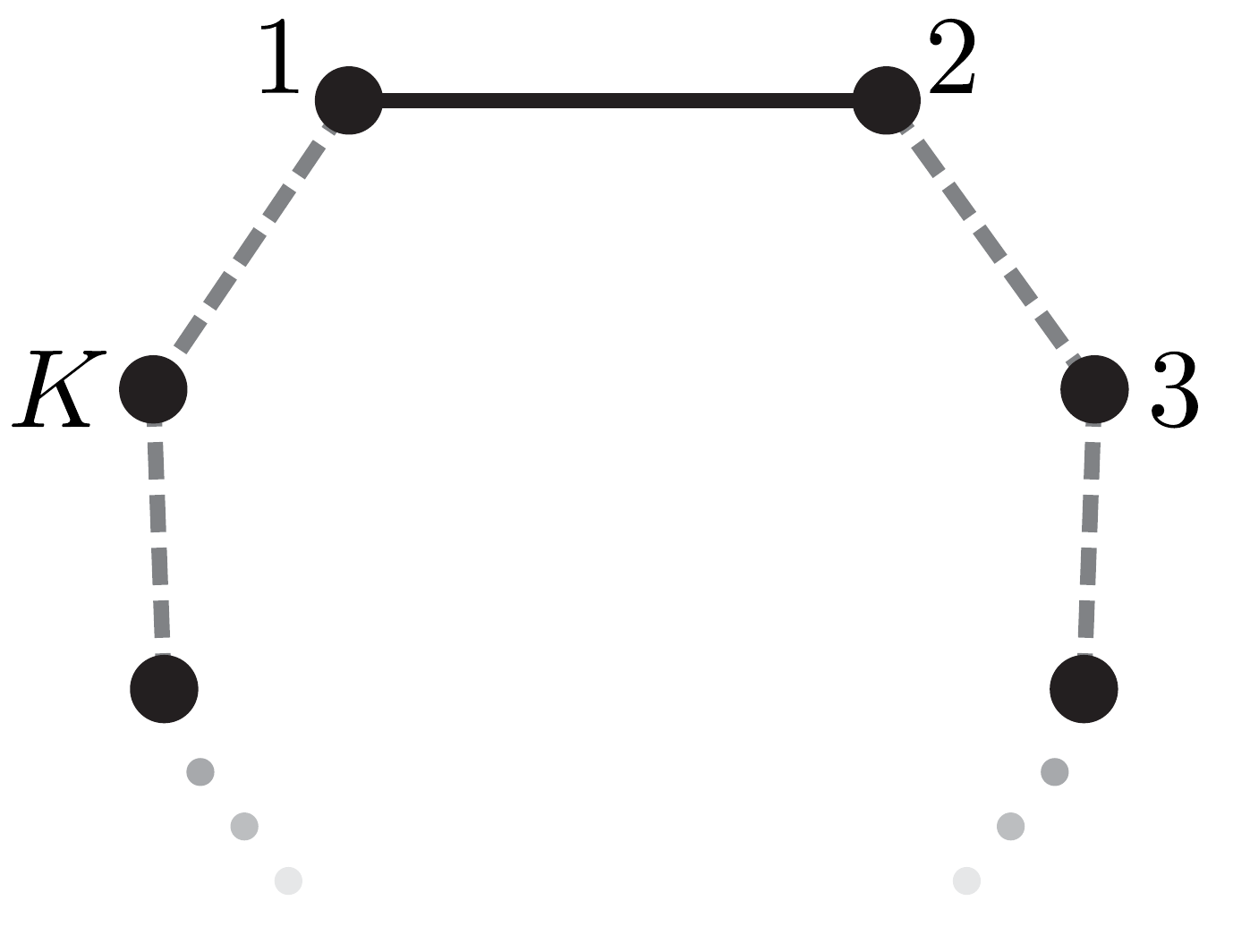}
\caption{\label{fig:unicycle} Illustration of a unicyclic network with $K$ states: Transitions between states $1$ and $2$ are observed, whereas all other transitions are hidden (depicted as gray dashed edges).}
\end{center}
\end{figure}

In addition to being a better estimate of the total entropy production, we have found that in unicyclic systems, as in Figure \ref{fig:unicycle}, the informed partial entropy production can saturate the hierarchy inequality \eqref{hierarchy} and provide the entire dissipation, which is our third main result.

In a unicyclic network, probability conservation requires that the steady-state current along every link is equal:
\begin{equation}
w_{i,i-1}\pi_{i-1}-w_{i-1,i}\pi_{i}=w_{i+1,i}\pi_{i}-w_{i,i+1}\pi_{i+1}.
\end{equation}
Hence, at stalling conditions, in addition to the vanishing of the current over the observed link, the currents over all the hidden transition are zero as well, and the system is actually at equilibrium.
Thus, at stalling, the ratio between the stalling probability distributions of states $1$ and $2$ simplifies to
\begin{equation}\label{stalling ratio in unicyle}
\frac{\pi_2^{\rm st}}{\pi_1^{\rm st}}=\frac{w_{23}\cdots w_{K1}}{w_{32}\cdots w_{1K}},
\end{equation}
which is a manifestation of detailed balance.
 Multiplying by the ratio of rates over the observed link $w_{12}/w_{21}$, gives the effective thermodynamic force of the informed partial entropy framework{
 \begin{equation}
 F^{\rm st}=\ln\frac{w_{12}\pi_2^{\rm st}}{w_{21}\pi_1^{\rm st}}=F^{\rm cycle},
 \end{equation}}
 which equals the cycle affinity of the entire unicylcic network \cite{Schnakenberg1976}.
 Hence,
\begin{equation}
\Sigma^{\rm ip}=J^\pi_{12}F^{\rm st}=J^\pi_{12}F^{\rm cycle}=\Sigma.
\end{equation} 

The example of the unicycle network clearly demonstrates an advantage of using the approach of Polettini and Esposito \cite{Polettini2017} in the case where only partial information is available and only a single link can be observed.
When the network contains no hidden cycles, extracting the stalling distribution of the two observed states is equivalent to having a complete information of the total entropy production in the system, rendering it the best inference strategy.

\section{Time-dependent partial entropy production}\label{sec:Time-dependent partial entropy production}
Having discussed some of the advantages of the informed partial entropy production \cite{Polettini2017}, we extend this approach to driven processes where rates are explicitly time dependent. 
Specifically,  we take the rates of the observed link to be time dependent through {an} external parameter protocol $X=\{x_t\}_{t=0}^T$, \emph{i.e.}, $w_{12}(t)\equiv w_{12}(x_t)$ and $w_{21}(t)\equiv w_{21}(x_t)$, whereas the rates of all the other transitions remain fixed.
In this case, the stalling distribution does not depend on time and the derivation of the fluctuation theorems for both the observed partial entropy production and the hidden entropy production carry through essentially unaltered.

To quote the result, let us introduce the {instantaneous} current $\phi_{ij}(t)$, counting the net number of jumps over each link as a function of time \cite{andrieux2007fluctuation},
\begin{equation}
\phi_{ij}(t)=\sum_{n=0}^{N-1}\delta(t-t_n)(\delta_{i,i_{n+1}}\delta_{j,i_n} - \delta_{j,i_{n+1}}\delta_{i,i_n}  ),
\end{equation}
where the system jumps from state $i_n$ to state $i_{n+1}$ at time $t_n$. Generalizing the definition of the partial entropy production along a trajectory to include the time dependency gives
\begin{equation}\label{FT_M_time_dependent}
\sigma^{\rm ip}=\ln\frac{\pi_{i_0}\pi^{\rm st}_{i_N}}{\pi_{i_N}\pi^{\rm st}_{i_{0}}}+\int_0^T dt\,  \phi_{12}(t)\ln\frac{w_{12}(t)\pi^{\rm st}_{2}}{w_{21}(t)\pi^{\rm st}_{1}}
\end{equation}
The corresponding fluctuation theorem is obtained by defining a time-dependent auxiliary process (cf. Eq.~\eqref{eq:def of Wip}):
\begin{equation}\label{Matteo auxiliary process time-dependent observed link}
 W_{ij}^{\rm ip}(t)=\left\{
\begin{array}{ll}
w_{ij}(t) & ij=12,21\\
w_{ji}\frac{\pi_i^{\rm st}}{\pi_j^{\rm st}} & i\neq j;\  ij\neq12,21 \\
-\lambda_{i}(t) & i=j=1,2 \\
-\lambda_{i} & i=j \neq1,2\\
\end{array}\right.
\end{equation}
The derivation is similar to the time-independent case.

The complementary entropy production $\sigma^{\rm ip,c}$ depends only on the rates of the unobserved link, and hence, does not change in this case.

\section{Numerical simulations}\label{sec:Numerical simulations}

\begin{figure}
\begin{center}
\includegraphics[scale=0.6]{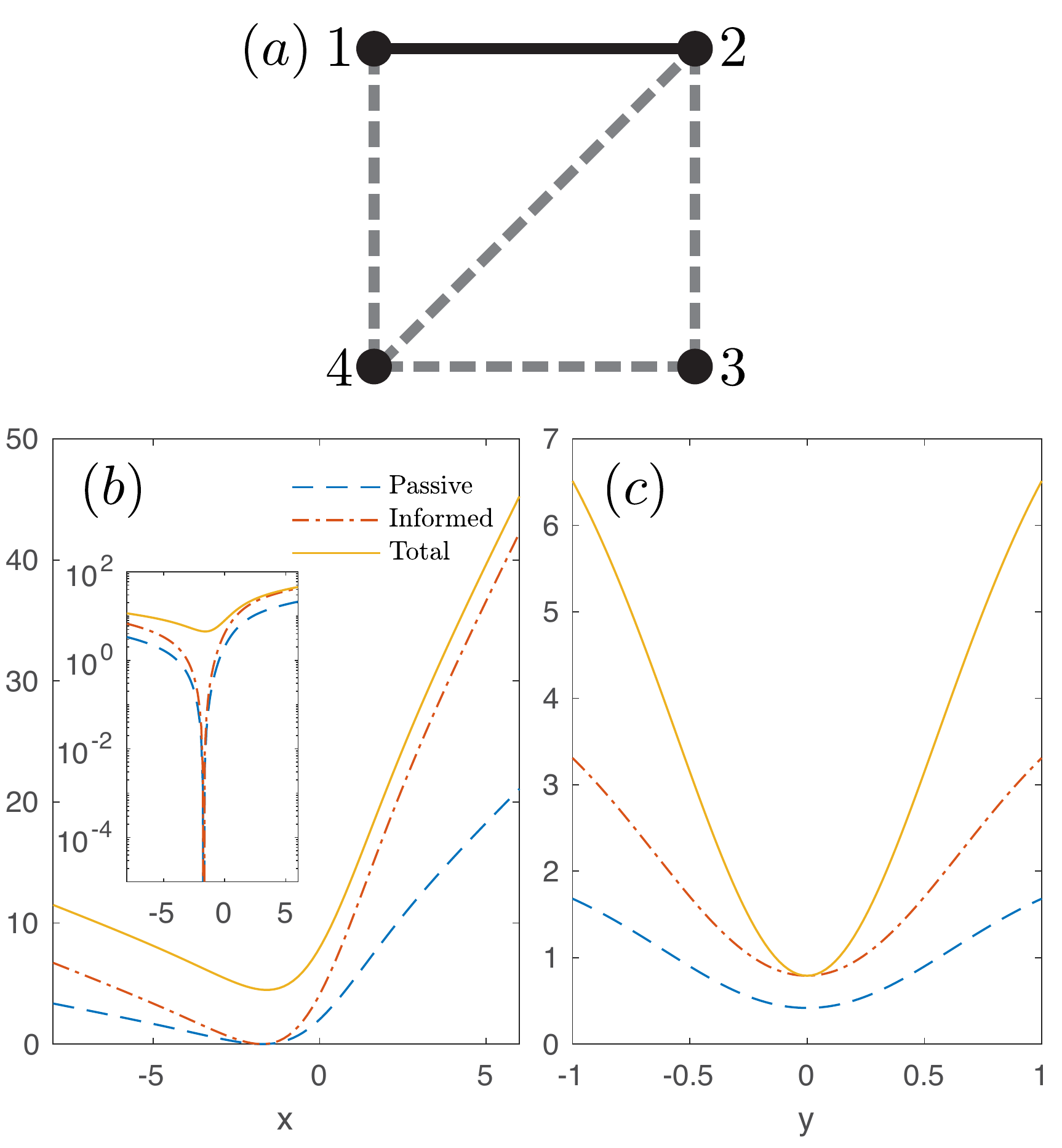}
\centering
\caption{ Entropy production rate with partial information: \emph{(a)} Network of states for a 4-state Markov process with generator $\bf W$ where link $1-2$ is observed. Passive (blue dashed curve), informed (red dotted-dashed curve), and total entropy production rates (solid yellow) with \emph{(b)} $w_{12}(x) = w_{12}e^{x}$, and $w_{21}(x) = w_{12}e^{-x}$. Inset: $y$-axis with logarithmic scale. \emph{(c)} $w_{24}(y) = w_{24}\sin^2(y)  $ and $w_{42}(y) = w_{42}\sin^2(y)$. Entries of the generator matrix are: $w_{12}=9$, $w_{13}=0$, $w_{14}=2$, $w_{21}=1$, $w_{23}=4$, $w_{24}=6$, $w_{31}=0$, $w_{32}=10$,  $w_{34}=5$,  $w_{41}=7$, $w_{42}=1$, $w_{43}=8$, where the diagonal elements {were} chosen to have zero-sum columns. {The control parameters are $y={\pi}/{2}$ in \emph{(b)} and $x=0$ in \emph{(c)}.}}\label{numerics}
\end{center}
\end{figure}

In order to illustrate our results, we randomly chose a single generator matrix $\bf W$ for a $4$-state system (Figure \ref{numerics}a), and numerically computed both the passive and informed partial entropy production rates, as well as the total entropy production rate for comparison. In our example, we observe the $1-2$ link with the rest of the network hidden.

The calculations were carried out for a set of such generator matrices, where we tuned the rates over the observed link with a control parameter $x$, $w_{12}(x) = w_{12}e^x$, and $w_{21}(x) = w_{21}e^{-x}$, where $w_{12}$ and $w_{21}$ are the original jump rates of $\bf W$. The range of values of the control parameter $x$ included the stalling force $x^{\rm st}$, which can be calculated in this case according to Eq. (\ref{eq:stalling rates and rates}),
\begin{equation}
x^{\rm st} = \frac{1}{2}\ln\frac{w_{21}\pi^{\rm st}_1}{w_{12}\pi^{\rm st}_2}.
\end{equation}
The results, depicted in Figure \ref{numerics}b, elucidate the entropy production hierarchy, and demonstrate that the informed partial entropy production rate is a better estimate of the total entropy production rate compared to the passive one. 
A clear limitation of both approaches is that an external observer cannot obtain a lower bound on the total entropy production at stalling conditions.

Further, in order to demonstrate that the informed partial entropy production can exactly predict the total entropy production for unicyclic networks, we used the same generator {$\bf W$ (with $x=0$)}, and tuned the hidden link $2-4$ according to $w_{24}(y) = w_{24}\sin^2(y)$ and $w_{42}(y) = w_{42}\sin^2(y)$, where $w_{24}$ and $w_{42}$ are the original entries of the generator matrix.  For $y=0$, the network becomes a single cycle. 
As can be seen in Figure~\ref{numerics}c, the informed partial entropy production rate converges to the total entropy production for $y=0$.

\section{Discussion}\label{sec:Discussion}

We have studied two notions of entropy production with partial information.
Their associated integral fluctuation theorems can be seen from one unifying perspective: each is obtained by comparing the system's dynamics to an auxiliary process that, in a manner of speaking, implements time-reversal on the unobserved part of the system.
Despite this similarity, the extra content embodied in the informed partial entropy production allows one to capture more of the underlying dissipation.
{The main challenge of this approach, however, is that the stalling force may be difficult to access in an experimental setup: isolating precise control of the transition rates only over the observed link may not be possible, as one might expect, for example, when monitoring a complex chemical reaction network within a living cell. }
{When it is applicable, however, the informed partial entropy production} offers a better estimate of the total entropy production rendering it a more useful inference tool; especially, for unicyclic networks where it captures all {of} the entropy production. 

Furthermore, in this work, we have extended the utility of the informed partial entropy production of Polettini and Esposito.
We have included the possibility of transient relaxation to the steady-state and driven nonautonomous processes, as well as developed a fluctuation theorem for the complementary entropy production in the unobserved subsystem.

To conclude, let us take a broader view of what has been discussed.
We have seen two different ways the total entropy production can be decomposed into two positive pieces that each verify a fluctuation theorem.
This is actually quite a remarkable property.
To appreciate this, let us try and decompose the total fluctuating entropy production in a similar manner by introducing an arbitrary auxiliary trajectory distribution ${\mathcal Q}$:
\begin{equation}
\sigma=\ln\frac{{\mathcal P}[\gamma]}{{\mathcal P}[\tilde \gamma]}=\ln\frac{{\mathcal P}[\gamma]}{{\mathcal Q}[\tilde \gamma]}+\ln\frac{{\mathcal Q}[\tilde\gamma]}{{\mathcal P}[\tilde \gamma]}.
\end{equation}
The first term as a ratio of trajectory probabilities with ${\mathcal P}$ in the numerator will satisfy a fluctuation theorem and will be positive on average: $\langle \ln({\mathcal P}/{\mathcal Q})\rangle_{\mathcal P}\ge0$, as it is the relative entropy between ${\mathcal P}$ and ${\mathcal Q}$.
The same cannot be said for the second term, because the original distribution ${\mathcal P}$ is in the denominator. 
However, the second term could be linked to an integral fluctuation theorem,  under a very special condition that 
\begin{equation}\label{eq:Rsymmetry}
\frac{{\mathcal Q}[\tilde\gamma]}{{\mathcal P}[\tilde \gamma]}=\frac{{\mathcal P}[\gamma]}{{\mathcal R}[\gamma^*]},
\end{equation}
for some possibly different trajectory distribution ${\mathcal R}$, with $\gamma^*$  either the original forward trajectory $\gamma$ or its time reverse $\tilde \gamma$.
For the passive partial entropy production, ${\mathcal R}$ turns out to be the symmetrical auxiliary process where the hidden part of the network becomes the observed part evaluated on the time-reverse trajectory \eqref{eq:ppPratio}.
For the informed partial entropy production, the auxiliary trajectory distribution remains unchanged, ${\mathcal R = \mathcal Q}$,  and is evaluated on the forward trajectory \eqref{eq:ipPratio}.
Identifying the general class of trajectory distributions for which the symmetry in \eqref{eq:Rsymmetry} holds, and thus allow a decomposition of the total entropy production into a pair of positive pieces that individually verify integral fluctuation theorems remains an open question.
However, understanding members of this class, as demonstrated in this work, can reveal deeper structure in the thermodynamics of nonequilibrium systems.

\ack
GB was funded by the James S. McDonnell Foundation for Scholar Grant No. 220020476. MP acknowledges support from the European Research Council (Project No. 681456).
TRG and JMH were supported by the Gordon and Betty Moore Foundation as Physics of Living Systems Fellows through Grant No. GBMF4513.

\section*{References}
\bibliographystyle{iopart-num}
\bibliography{FluctuationTheory.bib,Feedback.bib}

\clearpage

\appendix

\section{Full derivation of the partial entropy fluctuation theorems}\label{full_derivation_SL}
\subsection{Passive partial entropy production}
We start with the fluctuation theorem for the passive partial entropy production according to the approach of Shiraishi and Sagawa \cite{Shiraishi2014}. 
The transient fluctuation theorem is derived from the trajectory probabilities
\begin{equation}
\ln\frac{\mathcal P[\gamma]}{\bar{\mathcal P}^{\rm pp}[\tilde\gamma]}=\ln\frac{\pi_{i_0}}{\pi_{i_N}}+\sum_{n=0}^{N-1}\ln\frac{w_{i_{n+1},i_n}}{ \bar w_{i_{n},i_{n+1}}}
+\sum_{n=0}^{N}\ln\frac{e^{-\lambda_{i_n} \tau_{i_n}}}{e^{-\bar\lambda_{i_n} \tau_{i_n}}}.
\end{equation}
where we have assumed that the initial conditions of the auxiliary and original processes are sampled from the same steady-state distribution.
Next, we use the fact that 
\begin{equation}\label{eq:telescopic cancelation}
\sum_{n=0}^{N-1}\ln\frac{\pi_{i_n}}{\pi_{i_{n+1}}}=\ln\frac{\pi_{i_0}}{\pi_{i_{N}}},
\end{equation}
to get
\begin{equation}{\label{MEP4}}
\begin{split}
\ln\frac{\mathcal P[\gamma]}{\bar{\mathcal P}^{\rm pp}[\tilde\gamma]}&=
\sum_{n=0}^{N-1}\ln\frac{w_{i_{n+1},i_n}\pi_{i_n}}{ \bar w_{i_{n},i_{n+1}}\pi_{i_{n+1}}}-\sum_{n=0}^{N}{(\lambda_{i_n}-\bar\lambda_{i_n}) \tau_{i_n}}\\
&=\phi_{12}\ln\frac{w_{12}\pi_{2}}{w_{21}\pi_{1}}-\sum_{n=0}^{N}\left\{(\lambda_{1}-\bar\lambda_{1})\delta_{i_n,1} \tau_{i_n}+(\lambda_{2}-\bar\lambda_{2}) \delta_{i_n,2}\tau_{i_n}\right\},
\end{split}
\end{equation}
where $\phi_{12}$ is the total integrated current over the $1 - 2$ link, counting the net number of jumps from $2$ to $1$. Let us define $T_1$ to be the total time spent in state $1$ along the trajectory $T_1=\sum_{n=0}^{N}\delta_{i_n,1} \tau_{i_n}$, and similarly, $T_2$ is the total time spent in state $2$.
Then according to \eqref{single_link_duel_exit_time} and the fact that at steady state $\sum_{j\neq i}J^{\pi}_{ij}=0$, we have
\begin{equation}\label{diff_exit_rates}
\lambda_1-\bar\lambda_1=-\frac{1}{\pi_1}\sum_{j\neq1,2}J^{\pi}_{1j}=-\frac{1}{\pi_1}\underbrace{\sum_{j\neq1}J^{\pi}_{1j}}_{=0}+\frac{J^{\pi}_{12}}{\pi_1}=\frac{J^{\pi}_{12}}{\pi_1}.
\end{equation}
Similarly
\begin{equation}
\lambda_2-\bar\lambda_2=\frac{J^{\pi}_{21}}{\pi_2}.
\end{equation}
Allowing us to conclude that 
 \begin{equation}
\ln\frac{\mathcal P[\gamma]}{\bar{\mathcal P}^{\rm pp}[\tilde\gamma]}=\phi_{12}\ln\frac{w_{12}\pi_{2}}{w_{21}\pi_{1}}-\left(J^{\pi}_{12}\frac{T_1}{\pi_1}+J^{\pi}_{21}\frac{T_2}{\pi_2} \right),
\end{equation}
which completes the derivation. 

\subsection{Informed partial entropy production}
Let us now focus on the fluctuation theorem for the informed partial entropy production according to the approach of Polettini and Esposito \cite{Polettini2017}.
The transient fluctuation theorem is derived from the trajectory probabilities
\begin{equation}{\label{MEP1}}
\ln\frac{\mathcal P[\gamma]}{ \bar{\mathcal P}^{\rm ip}[\tilde\gamma] }=\ln\frac{\pi_{i_0}}{\pi_{i_N}}+\sum_{n=0}^{N-1}\ln\frac{w_{i_{n+1},i_n}}{ \bar w_{i_{n},i_{n+1}}},
\end{equation}
where we have assumed that the initial condition of the auxiliary process is sampled from the same distribution of the original process.
Next, similarly to the telescopicing cancelation in \eqref{eq:telescopic cancelation}, we use the fact that
\begin{equation}\label{eq:telescopic cancelation stalling}
\sum_{n=0}^{N-1}\ln\frac{\pi^{\rm st}_{i_n}}{\pi^{\rm st}_{i_{n+1}}}=\ln\frac{\pi^{\rm st}_{i_0}}{\pi^{\rm st}_{i_{N}}}
\end{equation}
to get
\begin{equation}{\label{MEP3}}
\ln\frac{\mathcal P[\gamma]}{ \bar{\mathcal P}^{\rm ip}[\tilde\gamma] }=\ln\frac{\pi_{i_0}\pi^{\rm st}_{i_{N}}}{\pi_{i_N}\pi^{\rm st}_{i_0}}+\sum_{n=0}^{N-1}\ln\frac{w_{i_{n+1},i_n}\pi^{\rm st}_{i_n}}{\bar w_{i_{n},i_{n+1}}\pi^{\rm st}_{i_{n+1}}}
=\ln\frac{\pi_{i_0}\pi^{\rm st}_{i_{N}}}{\pi_{i_N}\pi^{\rm st}_{i_0}}+\phi_{12}\ln\frac{w_{12}\pi^{\rm st}_{2}}{w_{21}\pi^{\rm st}_{1}},
\end{equation}
which completes the derivation. 

\section{Proof of the derivation of the stalling distribution}\label{Appendix:Stalling distribution}
The proof of \eqref{eq:stalling rates and rates} is based on the deletion-contraction formula \cite{Polettini2017}, where we denote by ${\bf W}_{(m_1,...,m_k|n_1,...,n_k)}$ the matrix obtained from ${\bf W}$ by removing the rows $\{m_1,...,m_k\}$ and columns $\{n_1,...,n_k\}$:
\begin{equation}\label{Proof_of_P_st_and_P_at_stalling}
\frac{\pi_1(x)}{\pi_2(x)}=\frac{w_{12}(x)\det {\bf W}_{(1,2|1,2)}+\det {\bf W}^{\rm st}_{(2|1)}}{w_{21}(x)\det {\bf W}_{(1,2|1,2)}+\det {\bf W}^{\rm st}_{(1|2)}}.
\end{equation}
At the stalling force $x^{\rm st}$, we thus have
\begin{equation}\label{eq: from which can define r}
\frac{\pi_1(x^{\rm st})}{\pi_2(x^{\rm st})}=\frac{w_{12}(x^{\rm st})+\frac{\det {\bf W}^{\rm st}_{(2|1)}}{\det {\bf W}_{(1,2|1,2)}}}{w_{21}(x^{\rm st})+\frac{\det {\bf W}^{\rm st}_{(1|2)}}{\det {\bf W}_{(1,2|1,2)}}}=\frac{\det {\bf W}^{\rm st}_{(2|1)}}{\det {\bf W}^{\rm st}_{(1|2)}},
\end{equation}
where for the second equality we used the fact that by definition, the current over the observed link is zero for $x^{\rm st}$:
\begin{equation}\label{current_at_stalling}
\frac{\pi_1(x^{\rm st})}{\pi_2(x^{\rm st})}=\frac{w_{12}(x^{\rm st})}{w_{21}(x^{\rm st})}.
\end{equation}
On the other hand, applying the formula in Eq. (\ref{Proof_of_P_st_and_P_at_stalling}) to the steady state distribution of the stalling matrix $\bf W^{\rm st}$ gives
\begin{equation}\label{stalling ratio}
\frac{\pi_1^{\rm st}}{\pi_2^{\rm st}}=\frac{\det {\bf W}^{\rm st}_{(2|1)}}{\det {\bf W}^{\rm st}_{(1|2)}},
\end{equation}
which proves Eq. (\ref{eq:stalling rates and rates}).

Let us note, that \eqref{eq: from which can define r} also defines the rates
\begin{equation}\label{eff rates}
r_{12}=\frac{\det {\bf W}^{\rm st}_{(2|1)}}{\det {\bf W}_{(1,2|1,2)}}, \qquad r_{21}=\frac{\det {\bf W}^{\rm st}_{(1|2)}}{\det {\bf W}_{(1,2|1,2)}}.
\end{equation}

\section{Full derivation of the passive hidden entropy production fluctuation theorem}\label{full_derivation_SM_unobserved}
We compare the natural logarithm of the forward trajectory generated by $\bf W$ and the time-reversed trajectory generated by $\bar{\bf W}^{\text{pp,c}}$ to obtain the passive hidden entropy production in accordance with the approach of Shiraishi and Sagawa \cite{Shiraishi2014},
\begin{equation}\label{ea:Ap hidden passive1}
\ln\frac{\mathcal{P}[\gamma]}{ \bar{\mathcal{P}}^{\rm pp,c}[\tilde\gamma] }=\ln\frac{\pi_{i_0}}{\pi_{i_N}}+\sum_{n=0}^{N-1}\ln\frac{w_{i_{n+1},i_n}}{ \bar w_{i_{n},i_{n+1}}}
+\sum_{n=0}^{N}\ln\frac{e^{-\lambda_{i_n} \tau_{i_n}}}{e^{-\bar\lambda_{i_n} \tau_{i_n}}}.
\end{equation}
We now use the fact that in the definition of $\bar{\bf W}^{\text{pp,c}}$ \eqref{hidden auxiliary process for single link}, only the rates corresponding to transitions over the $1-2$ link are ``twisted'', whereas the rest of the rates remain unaltered. 
Hence, the second term in the right hand side of \eqref{ea:Ap hidden passive1} can be split into two contributions
\begin{equation}
\begin{split}
\ln\frac{\mathcal{P}[\gamma]}{ \bar{\mathcal{P}}^{\rm pp,c}[\tilde\gamma] }&=\ln\frac{\pi_{i_0}}{\pi_{i_N}}+\phi_{12}\ln\frac{\pi_1}{\pi_2}+\sum_{\substack{i<j\\(i,j)\neq(1,2)}}\phi_{ij}\ln\frac{w_{ij}}{ w_{ji}}-\sum_{n=0}^{N}\left\{(\lambda_{1}-\bar\lambda_{1})\delta_{i_n,1} \tau_{i_n}+(\lambda_{2}-\bar\lambda_{2}) \delta_{i_n,2}\tau_{i_n}\right\}\\
&=\ln\frac{\pi_{i_0}}{\pi_{i_N}}+\phi_{12}\ln\frac{\pi_1}{\pi_2}+\sum_{\substack{i<j\\(i,j)\neq(1,2)}}\phi_{ij}\ln\frac{w_{ij}}{ w_{ji}}-\left\{T_1(\lambda_{1}-\bar\lambda_{1})+T_2(\lambda_{2}-\bar\lambda_{2}) \right\}\\
\end{split}
\end{equation}
where we have used the fact that only the escape rate of states $1$ and $2$ differ between $\bf W$ and $\bar{\bf W}^{\text{pp,c}}$, and the definition of $T_1$ and $T_2$ as the total time spent in the corresponding states along the trajectory. 
Plugging in the difference in escape rates \eqref{hidden single_link_duel_exit_time}, we find
\begin{equation}
\ln\frac{\mathcal{P}[\gamma]}{ \bar{\mathcal{P}}^{\rm pp,c}[\tilde\gamma] }=\ln\frac{\pi_{i_0}}{\pi_{i_N}}+\phi_{12}\ln\frac{\pi_1}{\pi_2}+\sum_{\substack{i<j\\(i,j)\neq(1,2)}}\phi_{ij}\ln\frac{w_{ij}}{ w_{ji}}+\left(J^{\pi}_{12}\frac{T_1}{\pi_1}+J^{\pi}_{21}\frac{T_2}{\pi_2} \right).
\end{equation}
We finally use the telescoping sum in \eqref{eq:telescopic cancelation} to combine the first two terms with the third,
\begin{equation}\label{eq:hidden passive3}
\ln\frac{\mathcal{P}[\gamma]}{ \bar{\mathcal{P}}^{\rm pp,c}[\tilde\gamma] }=\sum_{\substack{i<j\\(i,j)\neq(1,2)}}\phi_{ij}\ln\frac{w_{ij}\pi_j}{ w_{ji}\pi_i}+\left(J^{\pi}_{12}\frac{T_1}{\pi_1}+J^{\pi}_{21}\frac{T_2}{\pi_2} \right)=\sigma^{\rm pp,c}.
\end{equation}

\section{Full derivation of the entropy production decomposition}\label{full_derivation_decomposition}
\subsection{Passive partial entropy production}
Let us sum the contributions to the entropy production from both the observed link and the hidden part according to the passive partial entropy production approach,
\begin{equation}
\begin{split}
\sigma^{\rm pp}+ \sigma^{\rm pp,c} &=\phi_{12}\ln\frac{w_{12}\pi_{2}}{w_{21}\pi_{1}}-\left(J^{\pi}_{12}\frac{T_1}{\pi_1}+J^{\pi}_{21}\frac{T_2}{\pi_2} \right)+\\&\quad\quad\quad+\sum_{\substack{i<j\\(i,j)\neq(1,2)}}\phi_{ij}\ln\frac{w_{ij}\pi_j}{ w_{ji}\pi_i}+\left(J^{\pi}_{12}\frac{T_1}{\pi_1}+J^{\pi}_{21}\frac{T_2}{\pi_2} \right).
\end{split}
\end{equation}
We immediately see that the traffic terms (last terms in the first and second lines, respectively) cancel each other. This is exactly the reason for needing a different auxiliary $\bar{\bf W}^{\rm pp,c}$ process for the hidden dynamics -- to cancel the term resulting from the difference in escape rates in the original auxiliary process $\bar{\bf W}^{\rm pp}$. Combining the remaining terms we complete the derivation of the entropy production decomposition: 
\begin{equation}
\sigma^{\rm pp}+ \sigma^{\rm pp,c} =\sum_{i<j}\phi_{ij}\ln\frac{w_{ij}\pi_j}{ w_{ji}\pi_i}=\ln\frac{\pi_{i_0}}{\pi_{i_N}}+\sum_{i<j}\phi_{ij}\ln\frac{w_{ij}}{ w_{ji}}=\sigma.
\end{equation}

\subsection{Informed partial entropy production}
We sum the contributions of the informed partial entropy production of the observed and hidden parts:
\begin{equation}\label{Eq:Ap ip decomposition1}
\sigma^{\rm ip}+ \sigma^{\rm ip,c}=\ln\frac{\pi_{i_0}\pi^{\rm st}_{i_{N}}}{\pi_{i_N}\pi^{\rm st}_{i_0}}+\phi_{12}\ln\frac{w_{12}\pi^{\rm st}_{2}}{w_{21}\pi^{\rm st}_{1}}+\sum_{\substack{i<j\\(i,j)\neq(1,2)}}\phi_{ij}\ln\frac{w_{ij}\pi^{\rm st}_{j}}{ w_{ji}\pi^{\rm st}_{i}}.
\end{equation}
The second and third term on the right hand side of \eqref{Eq:Ap ip decomposition1} can be combined to a single sum, without the restriction on $(i,j)\neq(1,2)$:
\begin{equation}\label{Eq:Ap ip decomposition2}
\begin{split}
\sigma^{\rm ip}+ \sigma^{\rm ip,c}&=\ln\frac{\pi_{i_0}\pi^{\rm st}_{i_{N}}}{\pi_{i_N}\pi^{\rm st}_{i_0}}+\sum_{i<j}\phi_{ij}\ln\frac{w_{ij}\pi^{\rm st}_{j}}{ w_{ji}\pi^{\rm st}_{i}}.
\end{split}
\end{equation}
Now, the sum over ratios of stationary probabilities in the second term of \eqref{Eq:Ap ip decomposition2}, cancels telescopically (cf.~\eqref{eq:telescopic cancelation stalling}), except for initial and final contributions, which also cancel with the reciprocal ratio appearing in the first term:
\begin{equation}
\sigma^{\rm ip}+ \sigma^{\rm ip,c}=\ln\frac{\pi_{i_0}}{\pi_{i_N}}+\sum_{i<j}\phi_{ij}\ln\frac{w_{ij}}{ w_{ji}}=\sigma,
\end{equation}
which completes the proof.

\end{document}